\documentclass[%
 aip,
cp,  
 amsmath,amssymb,
 reprint,%
]{revtex4-2}

\usepackage{graphicx}
\usepackage{dcolumn}
\usepackage{bm}

\usepackage[utf8]{inputenc}
\usepackage[T1]{fontenc}
\usepackage{mathptmx} 

\begin{document}

\title{Constraining $f(R)$ Gravity Models with The Late-Time Cosmological Evolution}

\author{Ishfahani Rusyda}%
 \email{ishfahanirusyda@mail.ugm.ac.id}
\author{Romy H. S.  Budhi} 
 \email[Corresponding author: ]{romyhanang@ugm.ac.id}
\affiliation{
  Department of Physics, Universitas Gadjah Mada, Yogyakarta..
}
%

\date{\today} 

\begin{abstract}
The $f(R)$ Modified Gravity is a modification of Einstein's general theory of relativity, which aims to explain issues beyond The Standard Model of Cosmology such as dark energy and dark matter, which make up $68 \% $ and $27 \%$ of the energy density of the universe today, respectively. As a theory of gravitation that govern major dynamics on the large scale of the universe, an $ f(R)$ model should be able to explain the transition from a matter-dominated universe to a dark-energy-dominated universe to explain the recent accelerated expansion of the universe. Assuming that the density parameter of the radiation can be neglected during the transition from a matter-dominated universe to a dark-energy-dominated universe,  we find some fixed points regarding the dynamical stability of the density parameters of the model. The phase transition can be achieved if the $f(R)$ model can connect the fixed point $P_5$ (representing the matter-dominated era) to the fixed point $P_1$ (representing the dark energy-dominated era). The method to evaluate that state transition is called the Fixed-point analysis. 
In this study, we analyze the viability of $f(R)$ models proposed by Starobinsky, Hu-Sawicki, and Gogoi-Goswami regarding the phase transition from a matter-dominated universe to a dark-energy-dominated universe. It is shown that those models are viable by choosing some set of appropriate parameters. For example, in the Starobinsky and Hu-Sawicki models, the parameter $\mu$ can be chosen to correspond to the lower bound of $x_d= R_1/Rc$, where $R_1$ represents the de-Sitter point. Meanwhile, for the Gogoi-Guswami model, the same results can be achieved by taking  $\alpha$ and $\beta$ parameters satisfying the existence and stability conditions for the de-Sitter point. From these results, it can be concluded that those  $f(R)$ models allow such phase transitions of the universe to realize the late-time accelerated expansion.
\end{abstract}
\maketitle
\section{INTRODUCTION}

At its early age, the universe was very rapid and hot. Then it expanded and cooled until it became the present universe. According to the current research, cosmic inflation has succeeded in solving the horizon and flatness problem of the universe from the big bang theory  \citep{PhysRevD.23.347}. After inflation, the expansion of the universe can be divided into 3 phases \citep{baumann_2022}. The first phase is the radiation-dominated era when the universe's energy was dominated by relativistic particles. As it cooled, the relativistic particles lost their energy due to the expansion, annihilating, and decaying into non-relativistic particles, marking the beginning of the matter-dominated era. At the present time, the universe undergoes accelerating expansion in the dark-energy-dominated era which is supported by a lot of studies  \citep{article3,Perlmutter_1999, PhysRevD.60.081301, Spergel_2003, Riess_1998, riess1999bvri, PhysRevD.69.103501, PhysRevD.74.123507, Eisenstein_2005}.

The $\Lambda$CDM model is built on the assumption that Einstein's general theory of relativity is the correct theory to explain the universe on the cosmological scale. In the $\Lambda$CDM, the dark energy is represented by the cosmological constant $\Lambda$ and composes about 68\% energy of the universe. Another 27\% of the universe's energy is composed of dark matter, while the remaining 5\% consists of ordinary matter and radiation \citep{ryden_2016}. Despite its success, the $\Lambda$CDM has some flaws such as the fine-tuning and coincidence problem \citep{PERIVOLAROPOULOS2022101659}. Several solutions have been offered to solve this problem, one of which is by modifying the theory of gravity as it has been done in the $f(R)$ gravity \cite{DeFelice2010fRT,CLIFTON20121,10.1093/mnras/150.1.1}.

The $f(R)$ gravity, the main object of this research, is an effort to generalize the general theory of relativity by replacing the Ricci scalar $R$ in the Einstein-Hilbert action with a function of $R$. The $f(R)$ gravity has got a lot of attention due to its success in explaining inflation, dark matter, and dark energy just by modifying the gravitational sector of the Einstein-Hilbert action  \citep{DeFelice2010fRT}. In its development, there are some $f(R)$ models that have been proposed, such as the Starobinsky model \citep{Starobinsky2007} that could explain the cosmic inflation without modifying the particle sector, the Hu-Sawicki model \citep{PhysRevD.76.064004} that managed to explain the late-time acceleration phase of the universe, and the Gogoi-Goswami model \citep{Gogoi2020} that succeeded in explaining the propagation of gravitational wave and has passed the local astronomy constraint. To be considered viable, an $f(R)$ model has to account for the existence of the universe's phase transition, especially from the matter-dominated era to the dark-energy-dominated era. In this paper,  we analyzed the viability of $f(R)$ models proposed by Starobinsky,  Hu-Sawicki, and Gogoi-Guswami to explain the phase transition from the matter-dominated era to the dark-energy-dominated era by adopting so-called the fixed points analysis to investigated the dynamical stability of the density parameters of the model.

\section{\label{sec:level2}$f(R)$ COSMOLOGY}

The action of $f(R)$ gravity can be expressed in the following equation
\begin{equation}
    S= \int d^4x \left(\frac{\sqrt{-g} f(R)}{2\kappa} + L_M \right)
\end{equation}
where $\kappa^2 = 8 \pi G $  while $G $  is a bare gravitational constant, $f(R)$ is some arbitrary function of Ricci scalar $R$ derived from the metric $g_{\mu\nu}$ and $L_M$ is Lagrangian density of  the particle/matter sector. Here $g$ corresponds to the determinant of the metric. 
The field equation derived from the action by varying it concerning the metric $g_{\mu\nu}$ is
\begin{equation}\label{FEq}
    \kappa T_{\mu \nu} = FR_{\mu \nu} + g_{\mu \nu} \square  F - g_{\mu \nu} \nabla_\mu \nabla_\nu F -\frac{1}{2}g_{\mu \nu} f(R)
\end{equation}
where $F:= \partial f/\partial R$, $\nabla_\mu$ and $\square:= g^{\mu \nu} \nabla_\mu \nabla_\nu $ are the covariant derivative operator  and its the d'Alembertian operator associated to the metric $g_{\mu \nu}$, respectively,  and $T_{\mu \nu}$ is an energy-momentum tensor of matter. The trace of equation (\ref{FEq}) is given by
\begin{equation}\label{trace}
    3\square F(R) + F(R)R-2f(R)=\kappa^2 T
\end{equation}
The Einstein gravity  corresponds to $ f(R) = R$ and $F(R) = 1$, so that the term $f(R)$ in equation (\ref{trace}) disappears. In this case, we have the relation $R = -\kappa^2 T$, which means that the Ricci scalar R can be directly determined by the matter. In case $f(R) \neq R$, the term $F(R)$ does not vanish in equation (\ref{trace}), we can find a propagating scalar degree of freedom $ \phi:= F(R)$, which is called the scalaron,  where its dynamics can be obtained by solving the trace equation (\ref{trace}). There is a solution, so-called the de Sitter point, which corresponds to the vacuum condition ($T=0$) at which the Ricci scalar is constant $(\square F(R)=0)$. At the de Sitter point $R=R_1$, the above trace equation becomes
\begin{equation}
    \label{eks}
    F(R_1)R_1-2f(R_1)=0
\end{equation}
This equation becomes the condition for the existence of the de Sitter point. The future of the de Sitter solution may have decaying, growing, and oscillatory behavior.  To keep the stability of the de Sitter solution, an $F(R)$ should satisfy the following condition \citep{Gogoi2020, motohashi2011future}:
\begin{equation}
    \label{stab1}
    \frac{f_{,R} (R_1)}{f_{,RR}(R_1)}> R_1.
\end{equation}
where $f_{,R} =\partial f/\partial R= F(R)$ stands for  the first derivative of $f(R)$ with respect to $R$ and $f_{,RR} = \partial^2 f/\partial R^2$ is for  the second derivative of $f(R)$ with respect to $R$.

At the cosmological scale, we assume the universe is homogeneous,  isotropic, and expanding \citep{ryden_2016}. The homogeneous properties mean there is no special location in the universe, while the isotropic properties mean there is no special direction in the universe. Current investigations of the CMB measurements  (WMAP, BOOMERanG, and Planck for example) confirm that the universe is flat. Therefore, the geometry of the current universe can be represented through flat  Friedmann-Robertson-Walker (FLRW) space-time whose metric is given by \citep{Piattella2018}.
\begin{equation}
    ds^2= -dt^2 + a^2(t)\sum_{i=1}^3 dx_i^2
\end{equation}
where $a(t)$ is universe scale factor.  We also assume that the universe is filled with perfect fluids, with its energy-momentum tensor given by
\begin{equation}
    T^\mu_\nu = \mbox{diag}(-\rho,P,P,P )
\end{equation}
where $\rho$ denotes the energy density and $P$ are the pressure. These two quantities are related by the equation of state of matter $w:= P/\rho$.

After the cosmic inflation, the universe went through several phases of domination \citep{baumann_2022} :
\begin{enumerate}
    \item Radiation-dominated era. This era happened right after the inflation. In this era, the universe is dominated by relativistic particles. In this era, the equation of state is approximately given by $w\approx 1/3 $. 
    \item Matter-dominated era.
    This era happened after the universe cooled and the majority of relativistic particles become non-relativistic. In this era, its equation of state is $w\approx 0$
    \item Dark-energy-dominated era. After the matter-dominated era, the universe undergoes late time acceleration due to the domination of the dark energy, represented by the cosmological constant. In this era, we have $w\approx -1$
\end{enumerate}

An $f(R)$ model is viable if it satisfies the viability conditions:
\begin{equation}
    \label{frr}
    f_{,R} >0 , \quad f_{,RR} > 0, \quad  \mbox{for}  \quad R \geq R_0 
\end{equation}
 where  $R_0$ is Ricci scalar at the present time. The first condition is necessary to avoid the appearance of the ghost particle, a solution of the equation that exists in a phantom state ($w<-1$) and possesses negative kinetic energy, while the second condition is needed to prevent the squared mass $M^2$ of scalaron field (a scalar field that becomes the candidate of dark matter in $f(R)$) to be negative \citep{DeFelice2010fRT}. Aside from those two conditions, an $f(R)$ model has to be able to explain several conditions: producing a stable de Sitter solution, exhibiting the $\Lambda$CDM-like characteristic in the large curvature condition, having the value of $f(R=0)=0$ indicating the absence of cosmological constant at flat space-time, explaining late-time acceleration, accounting for the chameleon mechanism and passing the solar system test, and explaining universe phase transitions, which has become the main goal of this research \citep{Gogoi2020}.

\section{\label{sec:level3}FIXED-POINT ANALYSIS}
We use the fixed-point analysis which is written in \citep{PhysRevD.75.083504,cosdy}. First, consider a universe with a flat FLRW metric, filled with a perfect fluid composed of both matters with energy density $\rho_m$ and radiation with energy density $\rho_r$. The field equation from $f(R)$ action will take the form as 
\begin{equation}
    \label{3fh1}
    3FH^2 = (FR-f)/2 - 3H\dot{F} + \kappa^2 (\rho_m + \rho_r),
\end{equation}
\begin{equation}
    \label{fdd1}
    -2F\dot{H}=\ddot{F} - H \dot{F} + \kappa^2(\rho_m + (4/3)\rho_r)
\end{equation}
Using the following relations between $R,H,$ and $\dot{H}$ in the flat FLRW spacetime
\begin{equation}
    \label{R}
	R = 6(\dot{H} + 2H^2) \iff \dot{H} = \frac{R}{6} - 2H^2
\end{equation}
We can define these dynamic variables
\begin{equation}
    \label{252}
	\Omega_m =\frac{\rho_m}{3FH^2},  \quad x_4= \frac{\rho_r}{3FH^2} = \Omega_r  \quad x_3= \frac{R}{6H^2} \quad x_2= -\frac{f}{6FH^2} \quad x_1= -\frac{\dot{F}}{FH}
\end{equation}
that has relations with the density parameter of matter, radiation, and dark energy as followed
\begin{equation}
    \Omega_m= \frac{\kappa^2\rho_m}{3FH^2}=1-x_1-x_2-x_3-x_4, \quad \Omega_r = x_4, \quad \Omega_{DE}= x_1+x_2+x_3
\end{equation}
We define the number of e-foldings $N= \ln(a)$, so that $dN=d\ln(a)=H~dt$. From this definition, we can get the change of $x_1, x_2, x_3, x_4$ with respect to $N$ 
\begin{eqnarray}
    x_1' & =& \frac{d x_1}{d N} = - 1 - x_3 - 3x_2 + x_1^2 - x_1x_3 + x_4 \\
    x_2' &=& \frac{d x_2}{d N} = \frac{x_1x_3}{m} - x_2\left(2x_3 - 4 - x_1\right) \\
    x_3' &=& \frac{d x_3}{d N} = - \frac{x_1x_3}{m} - 2x_3\left(x_3 - 2\right) \\
    x_4' &=& \frac{d x_4}{d N} = -2x_3x_4 + x_1x_4 
\end{eqnarray}
while the value of $m$ and $r$ is defined by
\begin{equation}
    \label{m}
    m:= \frac{d\ln F}{d\ln R} = \frac{Rf,_{RR}}{f,_R} 
\end{equation}
\begin{equation}
    \label{r}
    r:= \frac{-d\ ln f}{d\ln R}=\frac{-Rf,_R}{f} = \frac{x_3}{x_2}
\end{equation}
For the case of flat FLRW metric in general relativity, the energy density and pressure of perfect fluid satisfy
\begin{eqnarray}
    \rho & =& \frac{3}{\kappa^2}H^2\nonumber \\
    p & =& -\frac{1}{\kappa^2}(3H^2+2\dot{H})
\end{eqnarray}
So the effective equation of state for the perfect fluid becomes
\begin{equation}
    w_{eff}= -1-2\dot{H}/(3H^2)= -(2 x_3-1)/3
\end{equation}
This equation can also be adapted for any $f(R)$ theory.

To study the transition from a matter-dominated era to a dark energy-dominated era, let's consider the case when the radiation component can be neglected $(x_4=0)$. We can find several fixed points $(x_1, x_2,x_3)$ that satisfy  $x_1'= x_2'=x_3'= x_4'= 0$ as following:
\begin{eqnarray}
	\label{p1} P_1 &:& (0,-1,2), \quad \mbox{with} \quad  \Omega_m=0  \quad \mbox{and} \quad w_{eff}=-1, \\
	\label{p2} P_2 &:& (-1,0,0) , \quad \mbox{with} \quad  \Omega_m=2  \quad \mbox{and} \quad w_{eff}=1/3, \\
	 \label{p3} P_3 &:& (1,0,0), \quad \mbox{with} \quad  \Omega_m=0  \quad \mbox{and} \quad w_{eff}=1/3,  \\
	 \label{p4} P_4 &:& (-4,5,0), \quad \mbox{with} \quad \Omega_m=0   \quad \mbox{and} \quad w_{eff}=1/3,\\
	 \label{p5} P_5 &:& \left ( \frac{3m}{1+m}, \frac{- 1+4m}{2(1+m)^2}, \frac{1+4m}{2(1+m)}\right), \quad \mbox{with} \quad \Omega_m=1 - \frac{m(7+10m)}{2(1+m)^2}   \quad \mbox{and} \quad w_{eff}=-\frac{m}{1+m},\\
	 \label{p6} P_6 &:& \left ( \frac{2(1-m)}{1+2m}, \frac{1-4m}{m(1+2m)}, - \frac{(1-4m)(1+m)}{m(1+2m)}\right), \quad \mbox{with} \quad \Omega_m=0   \quad \mbox{and} \quad w_{eff}=\frac{2-5m-6m^2}{3m(1+2m)}.
\end{eqnarray}
Therefore, the fixed point $P_1$ and $P_6$ represent the dark-energy-dominated era, while the matter-dominated era is represented by $P_5$ for the limit $m\rightarrow 0$. The dark-energy-dominated era has the equation of state $W_{eff}<-1/3$ dan $\Omega_m = 0$, indicating the assumed total absence of matter. Those requirements are satisfied by fixed points $P_1$ and $P_6$. The first fixed point $P_1$ is related to the de Sitter solution, the solution of Einstein's field equation in the flat universe, and the absence of any matter. This state corresponds to the dark-energy-dominated era and has $w_{eff}=-1$. 

The requirements for the existence and the stability of the de Sitter point can be written as a condition for $r$ and $m$, respectively
\begin{eqnarray}
    \label{rds} r=-2, \label{mds}\quad 0<m\leq 1
\end{eqnarray}
To represent the dark-energy-dominated era in $P_1$ point, the value of $m$ and $r$ have to satisfy those equations. There is another point that represents the dark-energy-dominated era, that is $P_6$ which is stable when the condition $m'_6<-1, (\sqrt{3}-1)/2 < m_6 < 1$ is satisfied. 

The matter-dominated era has the equation of state $w_{eff}=0$ and $\Omega_m = 1$. The fixed point $P_5$ that represents this era when $m\rightarrow 0$ is located along the line that satisfies:
\begin{equation}
    m(r)=-r-1. \quad 
    \label{mr1}
\end{equation}
Using equation (\ref{mr1}), we get $r=-1$ when $m\rightarrow 0$, so the matter-dominated era can be found around the point $(r,m)=(-1,0)$. From \cite{PhysRevD.75.083504}, for $m < 0$, the universe can't last long around $P_5$, so that value is not permitted. In order for $P_5$ point to be stable, the value of $m$ has to satisfy \cite{articleee}:
\begin{equation}
    \label{sp5}
    m(r < -1)>0 ,\quad  m'(r\leq-1)=\frac{dm}{dr} > -1, \quad m(r=-1)=0.
\end{equation}
From these conditions, we can say that the state with $m(r)= +0$ is needed for $f(R)$ to be viable if we want our theory to become $\Lambda$CDM-like in the matter-dominated era (large curvature region).

From this information, a viable $f(R)$ model from the perspective of explaining the transition of the universe can be divided into two classes :
\begin{enumerate}
    \item \textbf{Class A}: Model in this class connect $P_5 (r\simeq -1, m \simeq +0)$ point to $P_1 (r=-2,0<m\leq1)$ point.
    \item \textbf{Class B}: Model in this class connect $P_5 (r\simeq -1, m\simeq +0)$ point to $P_6 (m=-r-1,(\sqrt{3}-1)/2<m<1)$ point.
\end{enumerate}
In this paper, we will only discuss the viability of $f(R)$ from Class A.

\section{\label{sec:level4}RESULT AND DISCUSSION}
In this section, the result of the fixed-point analysis of Strarobinsky, Hu-Sawicki,  and Gogoi-Goswami's model will be discussed to explain the transition from matter dominated to a dark energy-dominated universe.

\subsection{\label{sec:level2a}Starobinsky Model}
The $f(R)$ Starobinsky model can be written as \citep{Starobinsky2007}.
\begin{equation}
    f(R)= R - \mu R_c \left[1-\left(1+\frac{R^2}{R_c^2}\right)^{-n}\right]
\end{equation}
with $n, \mu>0$ and $R_c$ is of the order of the presently observed effective cosmological constant. Here, $f(0)=0$ indicates that the cosmological constant vanishes in flat space-time, while $f(R\gg R_c)= R-\mu R_c$ indicates that the effective cosmological constant at high curvature is approximately given by  $\Lambda(\infty) \approx \mu R_c/2$.  
\begin{table}
\caption{\label{tabhu}The lower-bound of parameter $x_d$ dan $\mu$ in  Starobinsky model and  Hu-Sawicki model}
\begin{tabular}{ccccc}
\hline
 &\multicolumn{2}{c}{Starobinsky model}&\multicolumn{2}{c}{Hu-Sawicki model}\\
\hline
 $n$ &$x_d$ & $\mu$  &$x_d$ & $\mu$\\ \hline
		1 & $\geq$ 1.732050808  & $\geq$ 1.539600718  &  $\geq$ 1.732050808  & $\geq$ 1.539600718  \\ 
    3 & $\geq$ 1.041369159 & $\geq$ 0.7259417134  & $\geq$ 1.705308890 & $\geq$ 1.005145433   \\ 
    5 & $\geq$ 0.807987472 & $\geq$ 0.5323870496 & $\geq$  1.511940383 & $\geq$ 0.833814683  \\
		\hline
\end{tabular}
\end{table}


\begin{figure}[th!]
	\centering
	\includegraphics[width=0.9\linewidth]{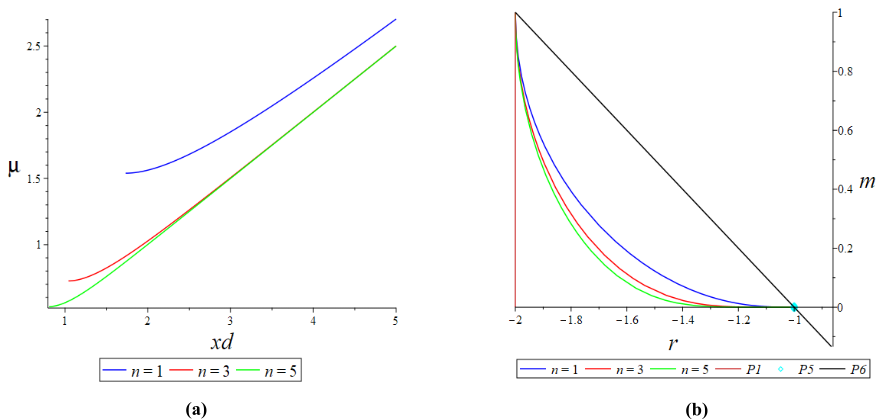}
	\caption{(a). Parameters $\mu$ as function of $x_d$ of the Starobinsky model for several value of $n$, (b). Plot of parameter $m$ as function of  $r$ in the Starobinsky model for several $n$ values. We take the lower bound of $\mu$ for each $n$ in this plot as listed in Table \ref{tabhu}. }
	\label{mrstra}
\end{figure}

First, we evaluate the existence and stability conditions at fixed-point $P_1$. Strarobinsky model satisfies the existence of de Sitter condition (\ref{eks}) for 
\begin{equation}
    R\left( 1- \frac{2 \mu \left(1 + \frac{R^2}{R_c^2} \right)^{-n}}{R_c \left( 1+ \frac{R^2}{R_c^2}\right)} \right) -2 R + 2 \mu R_c \left(1- \left( 1 + \frac{R^2}{R_c^2}\right)^{-n} \right)=0 
\end{equation}
calculated at Ricci scalar at de Sitter point $R=R_1$. By introducing $x_d=R_1/R_c>0$, the value of $\mu$ can be written as
\begin{equation}
    \mu=-{\frac {{\it x_d}\, \left( {{\it x_d}}^{2}+1 \right) }{ \left( 2+ \left( 2\,n+2 \right) {{\it x_d}}^{2} \right)  \left( {{\it x_d}}^{2}+1 \right) ^{-n}-2\,{{\it x_d}}^{2}-2}}
\end{equation}
The stability condition for the de Sitter point is analyzed by using equation (\ref{mds}). From this analysis, we obtain the values of $\mu$ and $x_d$ that satisfy de Sitter conditions and their stability. For example, assuming that $n$ is a positive integer, we obtain a lower bound for parameters $x_d$ and $\mu$ as listed in Table\ref{tabhu}. The plot of $\mu$ as a function of $x_d$ for various values of $n$ is illustrated in Fig.\ref{mrstra}.(a), while plot of parameter $m$ as function of  $r$ in Starobinsky model for several $n$ values. Fig.\ref{mrstra}.(b).
The figure shows that the greater $n$, the closer $\mu$ and $m$ to the former one. Thus behaviors of the model can be understood well by investigating several values of small  $n$.


Then, we analyze the stability condition for fixed-point $P_5$ by using equation (\ref{sp5}). For each value of $n$, at the large curvature region that represents matter dominated universe ($R\gg R_c$),  generally, this model has the values of $m \rightarrow 0$ and $r \rightarrow -1$ for any value of $\mu$. Viability condition at fixed-point $P_5$ that is $m'>-1$ is also satisfied because it has the value of $m'\rightarrow 0$ at  $R\gg R_c$.  
To visualize this transition, we chose  $\mu$  corresponds to the lower bound of $x_d$ and  representing the dynamics in  in $(m,r)$ plane. This plot can be seen in Fig.\ref{mrstra}. 

From Fig.\ref{mrstra}.(b), it can be seen that those graphs connect $P_5$ at $R\gg R_c$ to $(r,m)=(-2,1)$ which is de Sitter point $P_1$. Numerical investigation shows that taking the higher value of $\mu$ from its lower bound does not change the situation. Thus, as a conclusion, Starobinsky's model can describe the transition from a matter-dominated era to a dark-energy-dominated era.

\subsection{\label{sec:level2c}Hu-Sawicki Model}
The second model is proposed by Hu and Sawicki \citep{PhysRevD.76.064004}, which is written as
\begin{equation}
    f(R)= R-\mu R_c \frac{(R/R_c)^{2n}}{(R/R_c)^{2n} + 1}
\end{equation}
with $n, \mu, R_c >0$. As in Starobinsky's model, the parameter $\mu$ in Hu-Sawicki's model corresponds to the scale of the cosmological constant. For large curvature limit, we have  $f(R\gg R_c)= R-\mu R_c$, which indicates that the effective cosmological constant at high curvature is   $\Lambda(\infty) \approx \mu R_c/2$.
\begin{figure}[ht]
	\centering
	\includegraphics[width=0.9\linewidth]{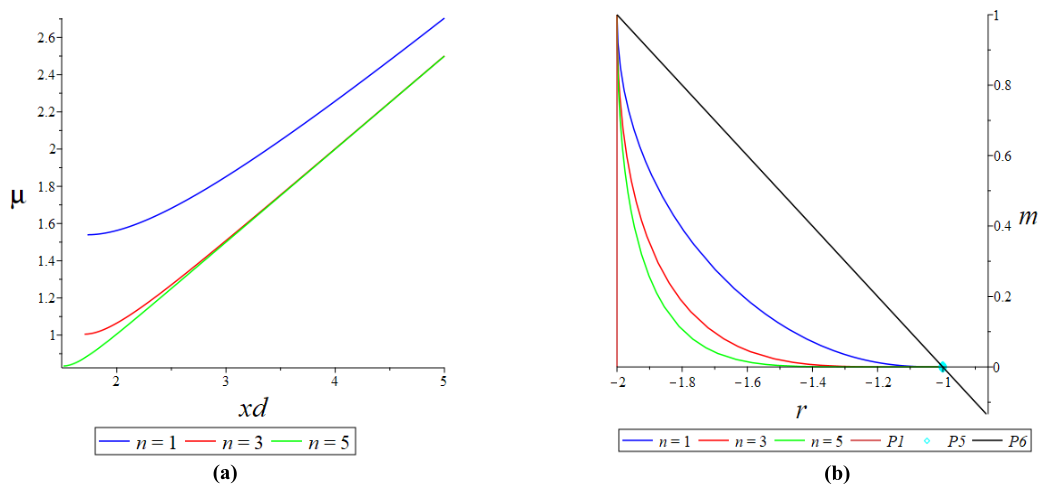}
	\caption{(a). Parameters $\mu$  as function of $x_d$ for Hu-Sawicki model, while (b). Plot $(m,r)$  as a function of $n$. We take parameter  $\mu$ as its lower bound as those in Table \ref{tabhu}.}
	\label{mrhuu}
\end{figure}

First, the viability of this model at de Sitter point had been evaluated and we derive that this model satisfies the existence condition in Equation (\ref{eks}) for 
\begin{equation}
    R\left( 1- \frac{2\mu R_c (R/R_c)^{2n} n}{R((R/R_c)^{2n} +1)} +  \frac{2\mu R_c ((R/R_c)^{2n})^2 n}{R((R/R_c)^{2n} +1)^2} \right) - 2R + \frac{2\mu R_c (R/R_c)^{2n}}{(R/R_c)^{2n} +1} = 0
\end{equation}
which is calculated at the de Sitter point $R=R_1$. As before, if we introduce $x_d= R_1/R_c$,  the value of $\mu$ can be written as
\begin{equation} \label{muhu}
    \mu = \frac{x_d((x_d^{2n})^2 + 2x_d^{2n}+1)}{2x_d^{2n}(x_d^{2n}-n+1)}.
\end{equation}
The stability condition of the de Sitter point given by equation (\ref{mds}) gives us the lower bound of parameter $x_d$ and $\mu$ as given in Table\ref{tabhu}. The table is listed for several values of integer $n$. Numerical study of the stability condition at de Sitter point in equation (\ref{mds})  and equation (\ref{muhu}) as a function of $x_d$ are illustrated in Fig.\ref{mrhuu}.(a).
As seen in the Starobinsky model, the larger $n$, the parameters $\mu$  tend to approach those of smaller  $n$. Thus, the model can be studied on some small $n$ values to represent its behavior.

Visualization of transition from $P1$ to $P5$ can be seen in $(m,r)$ plane in Fig.\ref{mrhuu} for each value of $n$. Here, we choose parameter $\mu$  as lower-bound of those listed in Table\ref{tabhu}. 
Figure \ref{mrhuu} shows that  the graph for each value of $n$ connect $P_5$ at  $R\gg R_c$ to de Sitter point $P_1$ at $(r,m)=(-2,1)$. Even if the figure is generated by using a lower bound of $\mu$ for each $n$, taking a higher value of $n$ generally yields a similar feature.

\subsection{\label{sec:level2d}Gogoi-Goswami Model}

The last $f(R)$ model we discuss is proposed by Gogoi-Goswami \citep{Gogoi2020} 
\begin{equation}
    \label{gogoi}
    f(R)= R- \frac{\alpha}{\pi} R_c \cot^{-1} \left(\frac{{R_c}^2}{R^2}\right)-\beta R_c \left[1-\exp\left(-\frac{R}{R_c}\right)\right]
\end{equation}
where $\alpha$ and $\beta$ are two dimensionless positive constants and $R_c$ is a parameter corresponding to the current Ricci scalar. At low curvature, we have $f(R\approx 0) = R$ indicating that this model reduced into Einstein gravity for low curvature region. Meanwhile, at large curvature region, we have $f(R\gg R_c) \approx R- R_c (\alpha+2\beta)/2$, which mimics the $\Lambda$CDM model in the large curvature region. 
\begin{figure}[th!]
	\centering
	\includegraphics[width=0.8\linewidth]{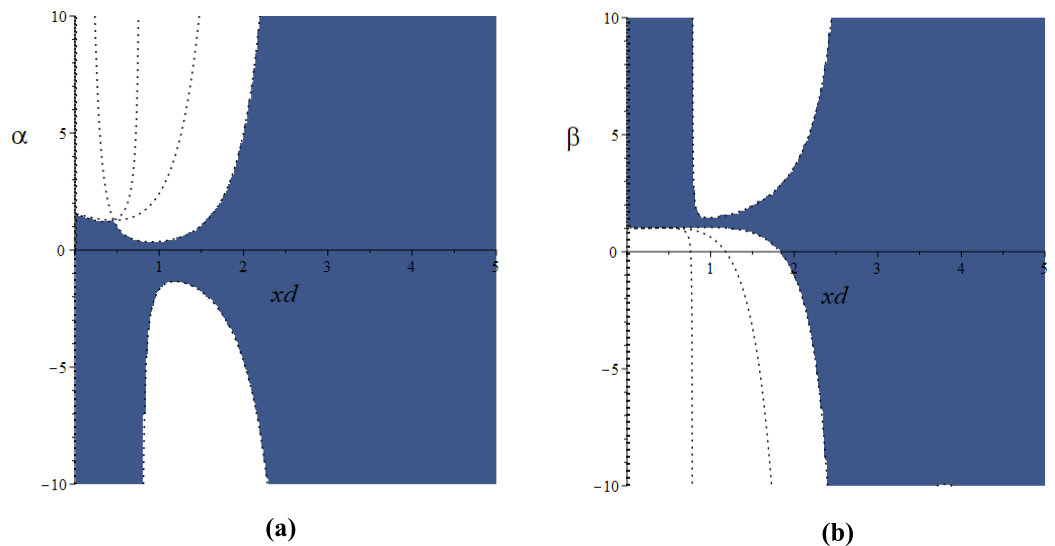}
	\caption{Viable interval of parameter $\alpha$ and $\beta$ as a function of $x_d$ in Gogoi-Goswami model as a result of stability condition of the de Sitter point.}
	\label{alpbetgoi}
\end{figure}

This model satisfies the existence of de Sitter condition in equation (\ref{eks}) for
\begin{equation}
    R \left(1-\frac{2 \alpha  R_{c}^{3}}{\pi R^{3} \left(\frac{R_{c}^{4}}{R^{4}}+1\right)}-\beta  \,{\mathrm e}^{-\frac{R}{R_{c}}}\right)-2 R + 2 \alpha\frac{  R_{c} }{\pi}  \mathrm{arccot}\! \left(\frac{R_{c}^{2}}{R^{2}}\right) +2 \beta  R_{c} \left(1-{\mathrm e}^{-\frac{R}{R_{c}}}\right)=0, 
\end{equation}
so that if we introduce $x_d=R_1/R_c$, where $R_1$ is Ricci scalar at de Sitter condition, the value of parameter $\beta$ can be written as
\begin{equation}
    \label{beta}
    \beta=\frac{\left(-2 \mathit{x_d}^{4} \alpha-2 \alpha\right) \arctan\! \left(\frac{1}{\mathit{x_d}^{2}}\right)+\mathit{x_d}^{4} \alpha \pi-\mathit{x_d}^{5} \pi-2 \mathit{x_d}^{2} \alpha+\alpha \pi-\mathit{x_d} \pi}{\left(\mathit{x_d}^{4}+1\right) \pi \left(-2+\left(\mathit{x_d}+2\right) {\mathrm e}^{-\mathit{x_d}}\right)}.
\end{equation}
 To evaluate its stability condition, we use equation (\ref{stab1}). This condition makes this model needs to satisfy equation (\ref{go1}) and (\ref{go2}):
\begin{equation}
    0 \leq \frac{-\pi \beta \left(\mathit{x_d}+1\right) \left(\mathit{x_d}^{4}+1\right)^{2} {\mathrm e}^{-\mathit{x_d}}+\pi \left(\mathit{x_d}^{4}+1\right)^{2}-8 \alpha \mathit{x_d}^{5}}{\pi \left(\mathit{x_d}^{4}+1\right)^{2}}
    \label{go1}
\end{equation}
\begin{equation}
    0< - \frac{(\pi \beta (x_d^4 +1)^2 e^{x_d} +6 \alpha x_d^4 -2 \alpha) x_d}{(\pi \beta (x_d^4 +1)e^{-x_d}-\pi x_d^4 + 2 \alpha x_d - \pi)(x_d^4 +1 )}
    \label{go2}
\end{equation}
Then, to analyze its stability at fixed-point $P_5$ we use equation (\ref{sp5}). From this analysis, we obtain that at large curvature region ($R\gg R_c$), this model has the value of $m \rightarrow 0$, $r \rightarrow -1$ and $m' \rightarrow 0$ for any value of $\alpha$ and $\beta$. Thus, condition (\ref{sp5}) is satisfied. 

\begin{figure}[th!]
	\centering
	\includegraphics[width=0.85\linewidth]{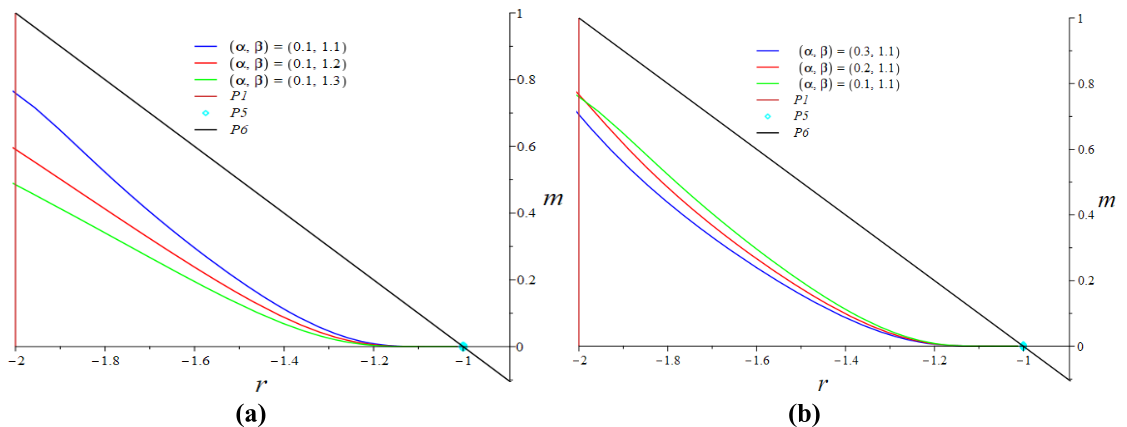}
	\caption{Parameters $m$ and $r$ in Gogoi-Goswami model. In (a) we take various $\beta$ for a fixed value of $\alpha$, meanwhile, in (b) we take various $\alpha$ for a fixed value of $\beta$.}
	\label{mrgo}
\end{figure}

From those analyses, we get the value of $\alpha$ and $\beta$ that satisfy the viability conditions at fixed-point $P_1$ and $P_5$. We show those values in Fig.\ref{alpbetgoi}. This graph shows that not all values of $\alpha$ and $\beta $ can be used for each  $x_d$. However, there are fairly narrow intervals in $\alpha$ and $\beta $ that possible to take for any value of $x_d$. Those interval is approximately about $(-1.215; 0.363)$ for $\alpha$, and  $(1.045; 1.386)$ for $\beta$. Taking a specific value of $\alpha,\beta$ in these intervals, we plot $(m,r)$ plane in Fig.\ref{mrgo}. 
To visualize this transition, we use the value of $\alpha$ and $\beta$ for $x_d=1$ to be represented in $(m,r)$ plane in Fig.\ref{mrgo}. 

The Fig.\ref{mrgo} shows that graphs for several variations of $\alpha$ and $\beta$ in the interval mentioned before,  connect $P_5$ at $(r,m)=(-1,0)$ to $(r,m)=(-2,1)$ which is de Sitter point $P_1$.Therefore, we can conclude that this model enables us to describe the transition from a matter-dominated era to a dark-energy-dominated era.

\section{CONCLUTION}
The transition from a matter-dominated universe to a dark-energy-dominated universe has been successfully evaluated by using fixed-point analysis. We obtain the values of parameters that satisfy both existence and stability conditions of relevant fixed-point for each $f(R)$ model. Those values are shown in Table \ref{tabhu} and visualized by using several graphs. 

\section{ACKNOWLEDGMENTS}

The authors would like to thank the Department of Physics, Universitas Gadjah Mada who have
facilitated us to do this research.


\bibliography{referensi-v4}
\end{document}